\documentclass[prl,floatfix,twocolumn,preprintnumbers,amsmath,amssymb,superscriptaddress]{revtex4}
\usepackage{color}

\usepackage{graphicx,psfrag}
\usepackage{hyperref}
\usepackage{verbatim}
\usepackage[utf8]{inputenc}

\newcommand{\be}{\begin{equation}}
\newcommand{\ee}{\end{equation}}
\newcommand{\bea}{\begin{eqnarray}}
\newcommand{\eea}{\end{eqnarray}}

\begin{document}
\title{A Note on Brane Inflation }
\author{Joseph P. Conlon}
\email{joseph.conlon@physics.ox.ac.uk}
\affiliation{Rudolf Peierls Centre for Theoretical Physics, Beecroft Building, Clarendon Laboratory, Parks Road, University of Oxford,
OX1 3PU, UK}
\begin{abstract}
This short note emphasises a potential tension between string models of inflation based on systems of branes and antibranes and the spectrum of strings in curved space, in particular the requirement that the leading Regge trajectory extends to the Planck scale allowing for the conventional string theory UV completion of gravity.
\end{abstract}
\maketitle

\textbf{Introduction: \,}
Inflation is the main paradigm for the formation of structure in the early universe and it is important to understand potential embeddings in string theory. In this respect, there are a wide variety of stringy inflationary scenarios with many different candidates for
the inflaton field (moduli, axions, brane modes....) -- a detailed review is \cite{BaumannMcAllister}.

In constructing scenarios, two clear, possibly distinct, goals exist. The first goal is to match observations and the second goal is to have parametric control of the scenario.
While parametrically controlled models may not be observationally viable, they are important for the existence claim that inflation can be realised in string theory. As with the related questions of moduli stabilisation and the construction of de Sitter vacua, it is important to know whether there are any \emph{principled and conceptual} difficulties with embedding inflation in string theory, or whether the problems are purely practical ones associated with the complexity of compactification scenarios. This is one aspect of the swampland program \cite{OoguriVafa}, for which \cite{EranReview} is a detailed review.

In this respect, one of the most popular stringy inflationary scenarios is brane inflation \cite{DvaliTye}. There are several reasons for its popularity. One is that it is rather generic, as branes are an unavoidable feature of perturbative string theory. It also has a simple interpretation, with the slope of the inflationary potential arising from the brane/antibrane interaction potential. Brane/antibrane inflation also offers novel phenomenology compared to field theory models of inflation, as the inflation field (the brane locus) simply disappears after inflation as the brane and antibrane annihilate.

This classical picture of branes is one that makes most sense at weak string coupling, when branes exist as solitonic non-perturbative objects. Once $g_s \sim 1$, there ceases to be a clear distinction between perturbative string modes and non-perturbative brane states. In general, a weak string coupling $g_s \ll 1$ is also normally assumed in moduli stabilisation scenarios, as it is helpful in ensuring control of the computations, by allowing the neglect of quantum loop corrections to the compactification data (or, at least, ensuring that only the leading such corrections are relevant).

It is true that the asymptotic limit of $g_s \to 0$ is a bad one from the perspective of matching observations. As in string theory the gauge couplings always in part inherit their strength from the string coupling, $\langle g_s \rangle$, the finite values of the Standard Model gauge couplings $\alpha_{SM} \sim \frac{1}{30}$ are unattainable for the case that $g_s \ll \frac{1}{30}$. However, such practical considerations are less relevant for the question of whether controlled string models of inflation are realisable in principle.

The question this note is concerned with is whether or not there are any principled difficulties in models of brane inflation with $g_s \ll 1$, and even more specifically with the asymptotic limit $g_s \lll 1$.

\textbf{Inflationary Brane/Antibrane Potentials: \,}
The literature on brane inflation models is large. This note is concerned with models in which a large part of the inflationary vacuum energy arises from the brane tension. This is true of brane/antibrane inflation and also of most variants of axion monodromy.

The bosonic part of the D-brane action is
\be
S = \frac{2 \pi}{l_s^k} \int {\rm d}^k {\rm x} \, e^{-\phi} \sqrt{g + 2 \pi \alpha' \mathcal{F}},
\ee
where $l_s = \frac{1}{m_s} = 2 \pi \sqrt{\alpha'}$ is the string length, $\phi$ is the string dilaton and $\mathcal{F}$ the internal gauge field on the brane.
For a brane that is space-filling but wraps an internal p-cycle $\Sigma$
of the compactification manifold, the vacuum energy of the brane is then
\be
V_{brane} = \frac{\textrm{Vol}_{\Sigma}}{g_s} \left( \frac{2 \pi}{l_s^4} \right).
\ee
Here $\textrm{Vol}_{\Sigma}$ is the volume of the internal cycle $\Sigma$ measured in units of $l_s^p$ (for a D3 brane that is
point-like in the internal space $p=0$ and $\textrm{Vol}_{\Sigma} = 0$). The vacuum energy associated to an antibrane is identical, as the two are instead distinguished by the sign of the Ramond-Ramond charge associated to the $C_p$ form under which the brane is charged.

This implies that in the doubly controlled limit of a compactification that is well described by classical geometry ($\textrm{Vol}_{\Sigma} > 1$) and has an asymptotically weak string coupling ($g_s \lll 1$), the energy associated to a brane-antibrane pair is $V_{D\bar{D}} \gg m_s^4$.
Note that this is also true for warped compactifications involving an anti-D3 brane at the bottom of a throat, except that here $m_s$ should now refer to the local string scale (either at the presence of the brane or the antibrane).

For such models, during the inflationary epoch the potential
always satisfies $V_{inf} \gg m_s^4$. For this not to occur requires a substantial cancellation between the brane tension
and other negative tension sources (such as orientifold planes) present in the compactification.
While this is possible for models which are fundamentally supersymmetric,
this is not the case for models involving explicit antibranes.

In any case, this note is restricted to the (large) class of models for which
the brane/antibrane tension contributes significantly to the overall vacuum energy, such that $V_{inflation} \gg m_s^4$.

\textbf{The Higuchi Bound: \,}
The relevance of this is that two recent papers \cite{EranDieter, 190702535} have pointed out a potential problem with an inflationary scale satisfying $V_{inflation} \gg m_s^4$. This arises from studying the string spectrum on de Sitter space, and in particular the spectrum for long rotating strings. As described in most detail in \cite{190702535}, the linear Regge trajectory
$$
E^2 \propto S,
$$
where $S$ is the spin and $E^2$ the squared mass, is modified for large spins as the string is longer and so feels the effect of the curvature of de Sitter space. This curvature results in there being both a maximum mass and a maximum spin along the Regge trajectory, which then curves back on itself to lower values of $E^2$ and $S$ (as illustrated in figure 3 of \cite{190702535}), with a maximal mass and spin of
\bea
E_{max} & = & 0.67 \frac{R}{\alpha'}, \\
S_{max} & = & 0.31 \frac{R^2}{\alpha'},
\eea
where $R$ is the de Sitter radius.

The implication of this is that for de Sitter radii $R \gtrsim M_P \alpha'$ (equivalently, a potential $V \lesssim \frac{1}{\alpha'^2} \sim m_s^4$), the maximal mass on the Regge trajectory is above the Planck scale. In this case
the conventional string theory ultraviolet completion of gravity can occur through excited string modes.

The converse of this is that for inflationary potentials with $V \gtrsim \frac{1}{\alpha'^2} \sim m_s^4$ the turnover in the Regge trajectory occurs before the Planck scale. In this case, the leading Regge trajectory has \emph{no} states at or near the Planck scale. In the extreme limit of a parametric separation $V \gg \frac{1}{\alpha'^2} \sim m_s^4$, the corresponding maximal energy along the leading Regge trajectory becomes
$E_{max} \ll M_P$  and the leading Regge trajectory contains no state near the Planck scale.

However, this parametric separation is exactly what occurs for brane inflation in the asymptotically controlled regime of $\textrm{Vol}_{\Sigma} > 1$ and $g_s \ll 1$; as $V \gg m_s^4$, the leading Regge trajectory turns over far below the Planck scale, causing a potential problem for the ordinary string theory ultraviolet completion of gravity. The region of parametric control therefore appears to lead to a new possible inconsistency.

Note that the use of warped throats does not affect this conclusion as the Regge trajectory comes from the spectrum of excitations in the non-compact spatial directions, which are unaffected by the warping.

\textbf{Caveats: \,}
For several reasons, this problem should be understood as a potential, rather than definite, problem. First, actual string compactifications involve
values of $g_s$ that are finite rather than arbitrary small. As $V \propto \frac{m_s^4}{g_s}$ implies $R \propto \sqrt{g_s}$, the
resulting numerical factor may be, for practical considerations, not that distinct from unity or any of the other
factors of $2 \pi$ entering the calculation. So even if there is a problem with UV completion of gravity for \emph{arbitrarily} small values of $g_s$, for finite values of $g_s$ the problem may not be significant.

Secondly, it is not clear that consistency requires the string theoretic UV completion of gravity to occur via the first (or leading) Regge trajectories. Although this is what happens in flat space, it is possible that for string theory on a de Sitter background the physics of the UV completion is different, without leading to any kind of inconsistency. A detailed resolution of this point would require the study of string scattering amplitudes in a curved background.

\textbf{Conclusions: \, }
The purpose of this short note is to emphasise, following the papers \cite{EranDieter, 190702535}, a potential problem
for brane inflation within the apparently parametrically controlled regime of weak couplings and large volumes. In the limit of $g_s \ll 1$ and cycle sizes ${\rm Vol}(\Sigma) \gg 1$,
the leading stringy Regge trajectory has a maximal energy $E_{max} \ll M_P$, invalidating the conventional string theoretic argument for an ultraviolet completion of gravity. For the reasons just outlined, this is a potential problem rather than a definite one; nonetheless, given the importance of the question of whether or not inflation and de Sitter vacua are realisable, it is important to scrutinise all such possible obstacles.

\bibliography{HiguchiRefs} % Produces the bibliography via BibTeX.
\end{document}